\documentstyle[aps,multicol,psfig]{revtex}

\begin{document}

\title{Solitons in quadratic nonlinear photonic crystals}
\author{Joel F.~Corney and Ole Bang}
\address{Department of Mathematical Modelling, 
Technical University of Denmark, Bldg.~321, 2800 Kgs.~Lyngby, Denmark}

\maketitle 

\begin{abstract}
We study solitons in one-dimensional quadratic nonlinear photonic crystals 
with modulation of both the linear and nonlinear susceptibilities.  
We derive averaged equations that include induced cubic nonlinearities 
and numerically find previously unknown soliton families.
The inclusion of the induced cubic terms enables us to show that solitons
still exist even when the {\em effective quadratic nonlinearity vanishes} 
and conventional theory predicts that there can be no soliton. 
We demonstrate that both bright and dark forms of these solitons
are stable under propagation.
\end{abstract}


\begin{multicols}{2}

\newlength{\figwidtha}
\newlength{\figwidthb}
\setlength{\figwidtha}{0.45\linewidth}
\setlength{\figwidthb}{0.95\linewidth}

The physics and applications of photonic band-gap (PBG) materials, or 
{\em photonic crystals}, have been active topics of research for more 
than a decade.
The theory of linear photonic crystals is now well understood, and many 
of their fundamental properties and technical applications have been 
characterized \cite{Sou96}.
The next important step in the application of photonic crystals is to 
create tunable PBGs.
Tunability is possible in linear photonic crystals through, e.g., the 
temperature dependence of the refractive index \cite{Leo_etal00} or the 
electro-optic effect \cite{BusJoh99}.
Ultrafast dynamical tunability of the PBG can be accomplished using
nonlinearity, as was first demonstrated with a constant Kerr nonlinearity 
\cite{LarHibMizSte90}.

Here we consider {\em quadratic nonlinear photonic crystals} (QNPCs) that have a linear grating (periodic dielectric constant) and/or a nonlinear 
grating (periodic second-order or $\chi^{(2)}$ susceptibility). 
QNPCs are of interest for all-optical components due to the fast and 
strong nonlinearity they can provide through the parametric cascading 
effect \cite{SteHagTor96}.  
The efficiency of the cascading process depends critically on the 
phase mismatch between the fundamental and second-harmonic (SH) waves,
but two powerful methods exist that use exactly a periodic photonic 
crystal structure to control the mismatch 
\cite{ArmBloDucPer62,TanBey73,JasArvLau86,FejMagJunBye92}. 
In one method, a QNPC with a linear Bragg grating is used to bend 
the dispersion curve near the PBG \cite{TanBey73,NPC_chi2_bragg}. 
However, the short period, which is of the order of the optical wavelength, can be 
inconvenient.
The second scheme, quasi-phase-matching (QPM), controls the phase mismatch 
\cite{ArmBloDucPer62,FejMagJunBye92} using a nonlinear grating with a
period equal to the comparatively long beat length (typically of the 
order of microns). 
QPM is also possible with linear gratings, but this is much less 
effective \cite{TanBey73,JasArvLau86}.

One of the spectacular manifestations of nonlinearity is the soliton, 
a self-localized entity that can propagate unchanged over long 
distances.
Homogeneous $\chi^{(2)}$ materials support solitons in all dimensions 
\cite{Tor98}, and gap solitons exist in QNPCs with a linear Bragg grating 
\cite{gap}.
In this Letter we focus on the open fundamental problem of whether solitons 
exist in 1D QNPCs with {\em both a linear and a nonlinear QPM grating}. 
Such a simultaneous linear grating is difficult to avoid when, for example, 
creating nonlinear QPM gratings in GaAs/AlAs semiconductors through 
quantum-well disordering \cite{Hel00}.

Solitons exist in 1D QNPCs with a nonlinear QPM grating \cite{ClaBanKiv97}, 
but a simultaneous linear grating can reduce the effective $\chi^{(2)}$
nonlinearity \cite{TanBey73,JasArvLau86,FejMagJunBye92}.
Thus the global existence of solitons in such QNPCs is nontrivial.
Building on previous findings that nonlinear QPM gratings induce 
cubic nonlinearities \cite{ClaBanKiv97}, we find soliton solutions 
that are stable under propagation.  
The induced cubic terms provide an elegant means of showing that, against 
intuition, {\em the QNPC supports stable bright and dark solitons even 
when there is no effective $\chi^{(2)}$ nonlinearity}.  
This is analogous to the existence of solitons in dispersion-managed 
fibres with no average dispersion \cite{DMsol}.

\begin{figure}
  \vspace{-8mm}
  \centerline{\hbox{\psfig{figure=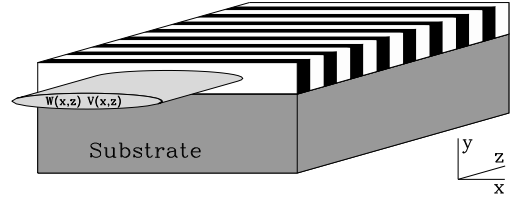,width=\figwidthb}}}
  \vspace{-5mm}
  \caption{Geometry of the 1D QNPC in the form of a $\chi^{(2)}$ slab 
  waveguide. Black and white domains indicate regions with different 
  dielectric constant and $\chi^{(2)}$ coefficient.} 
  \label{geometry} 
\end{figure}

We consider the interaction of a cw beam (carrier frequency $\omega$) 
with its SH, propagating in a lossless 1D QNPC under conditions for type I 
second-harmonic generation (SHG), as sketched in Fig.~\ref{geometry}.
We assume that the modulation of the refractive index is weak ($\Delta n_j(z)/\bar{n}_j$$\ll$1, where $n_j(z)$=$\bar{n}_j+\Delta n_j(z)$ and $j$ refers to the 
frequency $j\omega$), and we consider only gratings for forward QPM.
The grating period is then much longer than the optical period, in which 
case Bragg reflections can be neglected.  The evolution of the slowly 
varying beam envelopes is then described by
\cite{ArmBloDucPer62,MenSchTor94}
\begin{eqnarray}
  i\frac{\partial w}{\partial z} + \frac{1}{2}
  \frac{\partial^2 w}{\partial x^2} + \alpha_1(z) w + 
  \chi(z)w^*v {\rm e}^{i\beta z} & = & 0 ,\nonumber\\
  i\frac{\partial v}{\partial z} + \frac{1}{4}
  \frac{\partial^2 v}{\partial x^2} + 2\alpha_2(z) v + 
  \chi(z)w^2 {\rm e}^{-i\beta z} & = & 0, 
  \label{field_eqns}
\end{eqnarray}
where $w$=$w(x,z)$ and $v$=$v(x,z)$ are the envelope functions of the 
fundamental and SH, respectively.
The transverse and propagation coordinates $x$ and $z$ are in units of 
the input beam width $x_0$ and the diffraction length $L_d$=$k_1 x_0^2$, 
respectively. 
The parameter $\beta$=$\Delta k L_d$ is proportional to the mismatch 
$\Delta k$=$k_2$-$2k_1$, $k_j$=$j\omega\bar{n}_j/c$ being the average 
wavenumber. Thus $\beta$ is positive for normal dispersion and negative 
for anomalous dispersion.
The normalized refractive index grating is given by $\alpha_j(z)$=$L_d
\omega\Delta n_j(z)/c$ and the normalized nonlinear grating by 
$\chi(z)$=$L_d\omega d_{\rm eff}(z)/(\bar{n}_1 c)$, where 
$d_{\rm eff}$=$\chi^{(2)}/2$ is given in MKS units.
The model (\ref{field_eqns}) describes both temporal and spatial solitons
\cite{MenSchTor94}.

\begin{figure}
  \centerline{\hbox{\psfig{figure=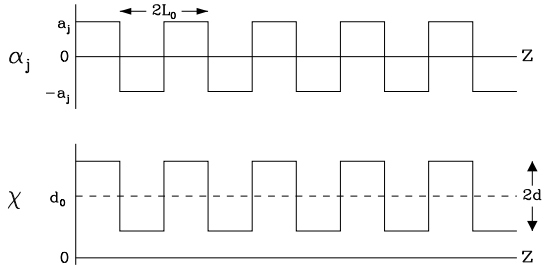,width=\figwidthb}}}
  \vspace{-5mm}
  \caption{Normalized linear and quadratic nonlinear gratings, 
  $\alpha_j(z)$ and $\chi(z)$, with period $2L_0$=$2\pi/|\kappa|$.} 
  \label{modulation}
\end{figure}

The aim is now to average Eqs.~(\ref{field_eqns}) and derive accurate
equations for the average field. To do so we focus on first order QPM 
using the conventional square gratings with 50\% duty cycle, shown in 
Fig.~\ref{modulation}.  We expand the grating functions in Fourier series:
\begin{equation}
  \label{FourGrating}
  \alpha_j(z) = a_j \sum_n g_n {\rm e}^{in\kappa z} , \;
  \chi(z)     = d_0 + d \sum_n g_n {\rm e}^{in\kappa z},
\end{equation}
where $g_n$=$2s/(i\pi n)$ for $n$ odd and $g_n$=0 for $n$ even, with $s = {\rm sign}(\kappa)$.
The gratings drive the system, which means that we may expand the 
envelope functions in Fourier series also:
\begin{equation}
  \label{FourField}
  w = \sum_n w_n(z,x) {\rm e}^{in\kappa z} , \quad
  v = \sum_n v_n(z,x) {\rm e}^{i(n\kappa - \tilde\beta)z} ,
\end{equation}
assuming that the coefficients $w_n(z,x)$ and $v_n(z,x)$ vary 
slowly in $z$ compared to $\exp(i\kappa z)$.
The residual mismatch $\tilde\beta$=$\beta-\kappa$ is ideally zero.  

Three physical length scales are in play: the diffraction length $L_d$, 
the coherence length $L_c$, and the grating domain length $L_0$. 
In normalized units $L_d$=1, $L_c$=$\pi/|\beta|$, and $L_0$=$\pi/|\kappa|$.
We assume a typical QPM grating with a domain length that is much shorter 
than the diffraction length, $L_0$$\ll$1. 
Furthermore, the grating is of good quality, with the domain length being 
close to the coherence length, $L_0$$\simeq$$L_c$, so the residual 
mismatch is small, $|\tilde\beta|$$\ll$$|\kappa|$.
In this case $|\kappa|$$\gg$1 and we can use perturbation theory with 
the small parameter $\epsilon$=$1/|\kappa| \ll 1$. 

Following the approach of Ref.~\cite{ClaBanKiv97}, we insert the Fourier
expansions (\ref{FourGrating}) and (\ref{FourField}) into the dynamical 
equations and assume the harmonics $w_{n\ne0}$ and $v_{n\ne0}$ to be of
order $\epsilon$. To lowest order ($\epsilon^1$), this gives the harmonics
\begin{eqnarray}
  w_{n\ne0} &=& (a_1g_nw_0 + dg_{n-1}w_0^*v_0)/(n\kappa), \nonumber\\
  v_{n\ne0} &=& (2a_2g_nv_0 + dg_{n+1}w_0^2)/(n\kappa).
\end{eqnarray}
Using these solutions, we obtain to first order $\epsilon$ the averaged 
equations for the DC components $w_0$ and $v_0$:
\begin{eqnarray}
  && i\frac{\partial w_0}{\partial z} +\frac{1}{2} 
     \frac{\partial^2 w_0}{\partial x^2} + \rho w_0^*v_0 +
     \gamma(|v_0|^2 - |w_0|^2)w_0 = 0, \nonumber\\
  && i\frac{\partial v_0}{\partial z} + \frac{1}{4} 
     \frac{\partial^2 v_0}{\partial x^2} + \tilde\beta v_0 + 
     \rho^* w_0^2 + 2\gamma|w_0|^2v_0 = 0.
  \label{average_eqns}
\end{eqnarray}
These equations also describe $m$th order QPM (where $\tilde\beta$=$\beta-m\kappa$
is ideally zero) and any other type of periodic grating, the parameters 
$\rho$ and $\gamma$ being simply given as sums over the Fourier coefficients 
of the grating \cite{ClaBanKiv97}. Incorporating time or the spatial $y$
coordinate is also straightforward.
For the square grating (\ref{FourGrating}),  $\rho$ and $\gamma$ can be explicitly 
calculated to
\begin{equation}
  \rho = i\frac{2d}{s\pi} + i\frac{4d_0(a_1 - a_2)}{s\pi\kappa}, \;  
  \gamma = \frac{d_0^2+d^2(1-8/\pi^2)}{\kappa}.  
\end{equation}

From Eqs.~(\ref{average_eqns}) follows the important result that {\em
cubic nonlinearities are induced in QNPCs} by nonlinear QPM gratings.
This cubic nonlinearity has the form of self- and cross-phase modulation 
(SPM and XPM), and is a result of non-phase-matched coupling between 
the wave at the main spatial frequency $\kappa$ and its higher harmonics. 
It is thus of a fundamentally different nature than the material 
Kerr nonlinearity, which is reflected in the fact that the SPM term 
is absent for the SH.

The averaged model (\ref{average_eqns}) is identical to the known model for 
nonlinear QPM gratings with no DC component ($a_j$=$d_0$=0), for which simulations 
confirmed that bright solitons had properties that were not predicted by the 
conventional model with only quadratic terms, but were accurately described 
with the inclusion of the cubic terms \cite{ClaBanKiv97}.
It was further shown that the induced cubic nonlinearity affects 
the phase modulation of cw waves, enabling efficient switching 
\cite{KobLedBanKiv98}, and that its strength can be increased
by modulating the grating \cite{BanClaChrTor99}. 

For the more general QNPCs considered here, the induced cubic nonlinearity 
depends on both the DC part and the modulation part of the nonlinear 
grating, but is independent of the linear grating. 
The cubic terms may lead to either a focusing or a defocusing effect, 
depending on the relative intensity of the fields and the sign of the 
phase mismatch $\beta$, since ${\rm sign}(\kappa)= {\rm sign}(\beta)$.  
The strength of the effective $\chi^{(2)}$ nonlinearity depends on the 
difference in the linear grating strengths at the fundamental and SH 
frequencies and on the DC component of the nonlinear grating.  
We thus recover the well-known effect that the interplay between the 
linear and nonlinear gratings can increase or decrease the effective 
$\chi^{(2)}$ nonlinearity, depending on the physical situation 
\cite{TanBey73,JasArvLau86}. 

The averaged model (\ref{average_eqns}) has stationary, localised soliton
solutions of the form $w_0(x,z)$=${\rm e}^{i \lambda z}\tilde w(x)/|\rho|$ 
and $v_0(x,z)$=${\rm e}^{2i\lambda z}\tilde v(x)/\rho$, which obey the 
equations
\begin{eqnarray}
  & & \frac{1}{2} \frac{\partial^2\tilde w}{\partial x^2} - \lambda\tilde w
      + \tilde w \tilde v + \tilde\gamma(\tilde v^2 - \tilde w^2)\tilde w 
      = 0, \nonumber\\
  & & \frac{1}{4}\frac{\partial^2 \tilde v}{\partial x^2} + 
      (\tilde\beta-2\lambda) \tilde v + \tilde w^2 + 
      2\tilde\gamma \tilde w^2\tilde v = 0,
  \label{soliton_eqns}
\end{eqnarray}
where $\tilde\gamma$=$\gamma/|\rho|^2$ depends only on $\kappa$, $a_1-a_2$,
and $d_0/d$.  
The slowly varying approximation gives valid solutions when the soliton period is longer than the grating period, i.e.~when the 
soliton parameter $\lambda$ is small, $|\lambda|\ll |\kappa|$.  

\begin{figure}
  \centerline{\hbox{
  \psfig{figure=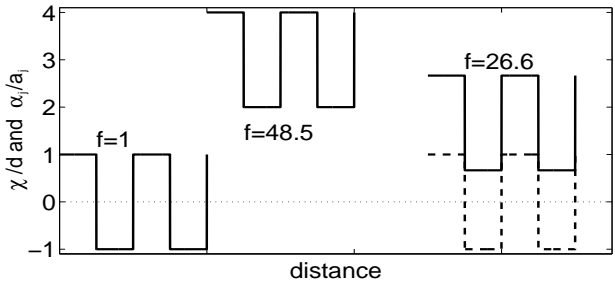,width=\figwidthb,height=\figwidtha}}}
  \caption{QNPCs with the same value of $\tilde\gamma$. The linear 
  (nonlinear) grating is shown with a dashed (solid) curve.}
  \label{phys_grating}
\end{figure}

It is important to stress that Eqs.~(\ref{soliton_eqns}) cover a much more
general situation than in \cite{ClaBanKiv97}, which only considered a nonlinear grating.  
A given value of the parameter $\tilde\gamma$ represents {\em a range of 
physical situations with different combinations of linear and nonlinear 
gratings}.  
In Fig.~\ref{phys_grating} we illustrate representative combinations 
for exact phase matching ($\tilde\beta$=0) that all give the same 
value of $\tilde\gamma$.

The first, simple case ($f$=1) is typical for domain inversion in 
ferroelectric materials, such as LiNbO$_3$. It has only a nonlinear 
grating that flips from positive to negative with no DC component, $d_0$=0.  
The second case has a DC level of $\chi^{(2)}$ nonlinearity, $d_0/d$=3,
but no linear grating, corresponding to the nonlinear part of the 
LiNbO$_3$/H:LiNbO$_3$ structure reported in \cite{Pet96}.  
The third case is the GaAs/GaAlAs structure reported in \cite{Pet96}, which
has a nonlinear grating with $d_0/d$=5/3 and a linear grating with 
$(a_1-a_2)/\beta$=$-0.07$.  
For the second and third cases to give the same $\tilde\gamma$ as the 
first, the grating wavenumber $\kappa$ must be multiplied by $f$=48.5 
and $f$=26.6, respectively. 
Physically this can be done by changing 
the input beam width $x_0$, thus maintaining $\tilde\beta$=0.

\begin{figure}
  \centerline{\hbox{
  \psfig{figure=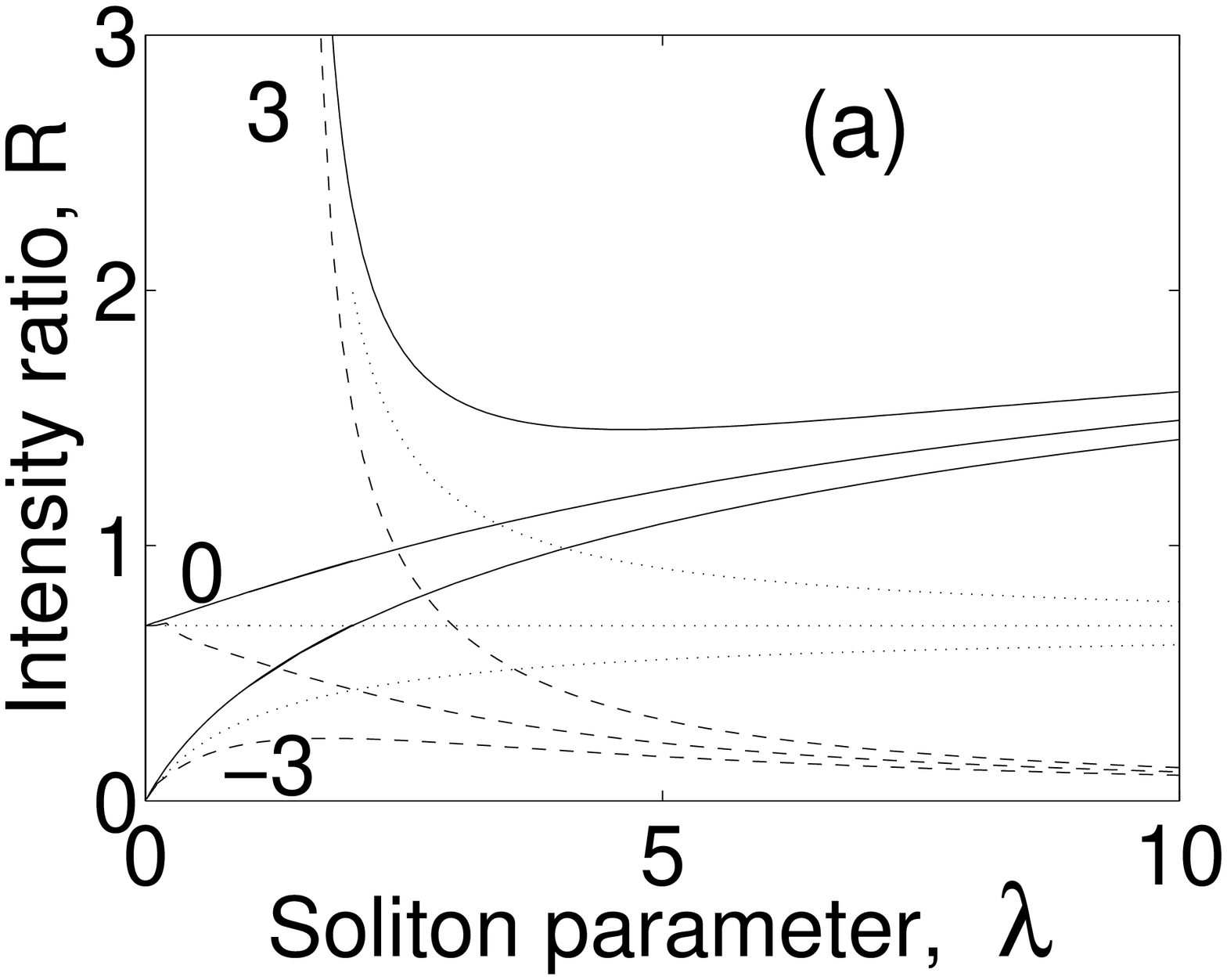,width=\figwidtha}
  \psfig{figure=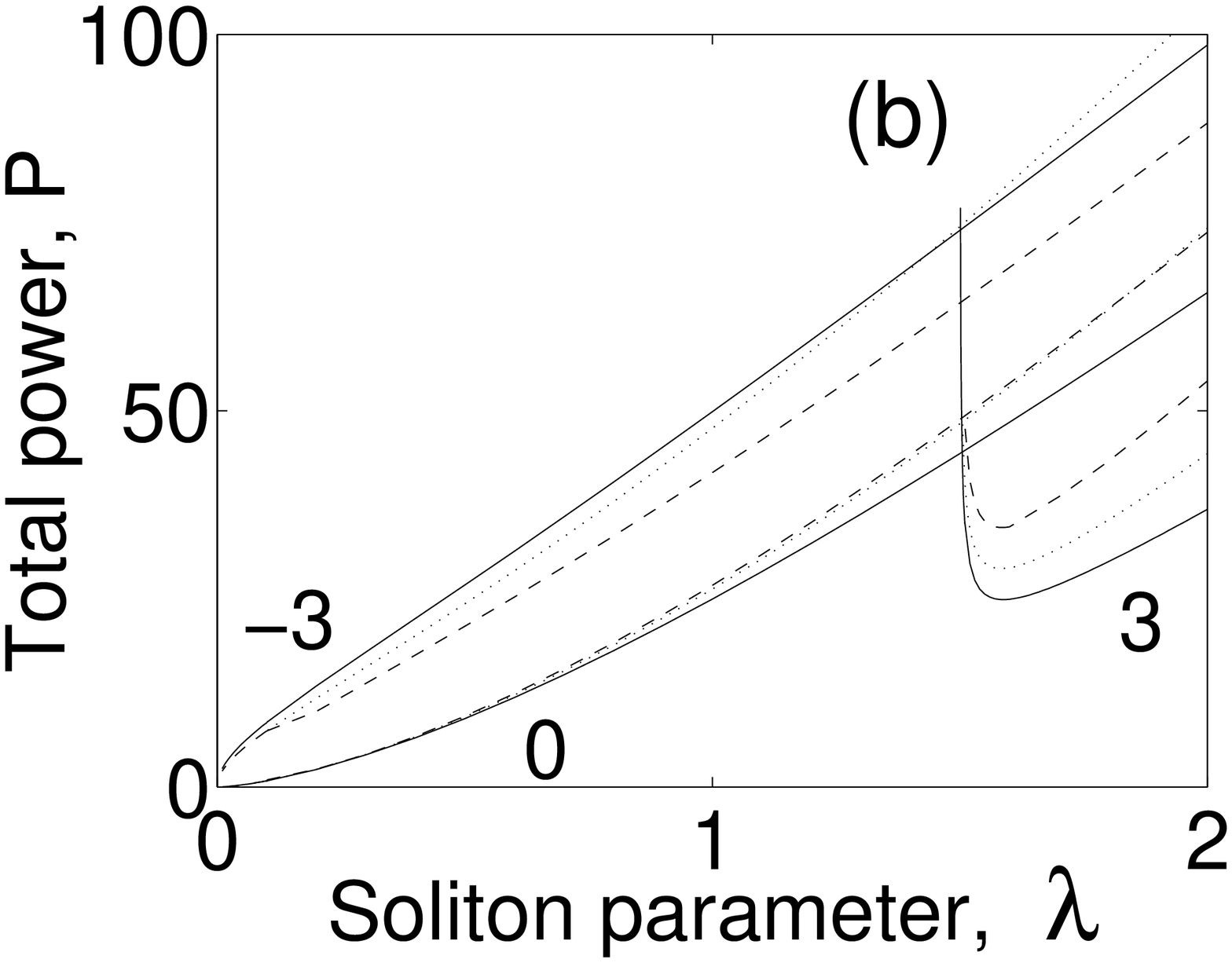,width=\figwidtha}}}
  \caption{Soliton properties versus the internal parameter $\lambda$ for 
  $\tilde\gamma$=0.02 (solid), $\tilde\gamma$=$-0.02$ (dashed), and 
  $\tilde\gamma$=0 (dotted), and three values of the residual mismatch 
  $\tilde\beta$. (a) Ratio of peak intensities $R$=$\tilde v^2(0)/\tilde 
  w^2(0)$. (b) Total power $P$.} 
  \label{sol_families}
\end{figure}

We have numerically found the bright soliton solutions of Eqs.~(\ref{soliton_eqns}) using a standard relaxation technique.
Bright soliton properties were investigated in \cite{ClaBanKiv97},
but not for normal dispersion, $\tilde\gamma>0$.
In Fig.~\ref{sol_families} we present the properties for normal 
($\tilde\gamma$=0.02) and anomalous ($\tilde\gamma$=$-0.02$) dispersion, 
together with the zeroth-order solution ($\tilde\gamma$=0).
The ratio $R=\tilde v^2(0)/\tilde w^2(0)$ of peak intensities, shown 
in Fig.~\ref{sol_families}(a), confirms that the zeroth-order approximation 
becomes increasingly inaccurate for large $\lambda$.  
Also, for a given $\tilde\gamma$, $R$ approaches the same limiting value as 
$\lambda$ increases, regardless of the value of $\tilde\beta$.  
In this limit the SH is stronger than the fundamental for $\tilde\gamma>0$
($R>1$) and much weaker for $\tilde\gamma<0$ ($R\simeq0$).  
The total power $P$=$\int_{-\infty}^\infty(\tilde v^2 + \tilde w^2)dx$ is 
shown in Fig.~\ref{sol_families}(b).  For $\beta > 0$ this reveals the interesting property that the
power threshold for existence, decreases for $\tilde\gamma>0$ 
and increases for $\tilde\gamma<0$, as compared to the zeroth-order value.
 
\begin{figure}
  \centerline{\hbox{\psfig{figure=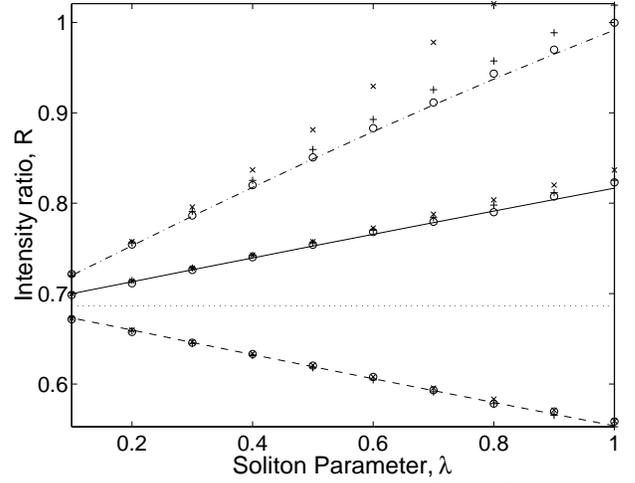,width=\figwidthb}}}
  \caption{Ratio of peak intensities $R$=$\tilde v^2(0)/\tilde w^2(0)$ versus 
   $\lambda$ for $\kappa$=$10f$ (continuous), $\kappa$=$4f$ (dot-dashed), and 
   $\kappa$=$-10f$ (dashed), and for the zeroth-order solution (dotted). 
   The average of the propagating solitons are shown with $f$=1 (crosses), 
   $f$=48.5 (plusses), and $f$=26.6 (circles). $\tilde\beta$=0.} 
  \label{compare}
\end{figure}

The bright soliton solutions of the average model (\ref{average_eqns}) were 
tested for the three QNPCs of Fig.~\ref{phys_grating} by mapping them back 
to the variables $w$ and $v$, and using them as initial conditions in 
simulations of the field Eqs.~(\ref{field_eqns}).  
The evolution consists of small, regular oscillations superimposed on the slow 
average beam.  
Properties of the propagating solitons were calculated by averaging over
an integer number of grating periods and were then compared with the
predictions of the average model.  
Figure \ref{compare} displays the ratio of peak intensities versus $\lambda$
for exact phase matching, $\tilde\beta$=0, and reveals that, for both 
anomalous and normal dispersion, the solutions of the average model are
accurate for small $\lambda$ and large $|\kappa|$, as expected.  
Even when $|\kappa|$=4 the first-order solutions provide a much better 
fit than the zeroth-order solutions.  
Our analysis thus shows that bright solitons exist and propagate stably
in QNPCs with many types and combinations of linear and nonlinear QPM gratings.

However, this is provided one is careful and does not eliminate the effective 
$\chi^{(2)}$ nonlinearity by using a grating with $\kappa$=$2(a_2-a_1)d_0/d$. 
This could happen in realistic QNPCs without violating the assumption
$|\kappa|\gg1$.
For the LiNbO$_3$ and GaP/AIP structures given in \cite{Pet96} at 
phase matching ($\kappa$=$\beta$), the presence of the linear grating 
changes the effective $\chi^{(2)}$ nonlinearity by a factor of 
${\cal F}$=$1+2d_0(a_1-a_2)/(d\kappa)$=1.4 and 0.3, respectively.  
The linear grating thus adds constructively in the LiNb0$_3$ structure and 
destructively in GaP/AIP.  In fact, modifying the nonlinear grating in the GaP/AIP structure slightly to 
$\chi^{(2)}_a$=40pm/V (max.) and $\chi^{(2)}_b$=19pm/V (min.) eliminates the effective 
$\chi^{(2)}$ nonlinearity entirely.

Conventional average models \cite{TanBey73,JasArvLau86,FejMagJunBye92} 
would predict that no soliton could exist with no nonlinearity, $\rho$=0.
However, in the model (\ref{average_eqns}) the induced cubic nonlinearity 
predicts that solitons should still exist as solutions of nonlinear 
Schr\"odinger equations.
In the case when the SH is 
strong ($v_0/w_0=\sqrt{5}\exp[i(\tilde\beta-\lambda/2)z]$), a family of bright solitons, $w_0$=$\sqrt{\lambda/(2\gamma)}\,{\rm sech}
(\sqrt{2\lambda}x){\rm e}^{i\lambda z}$,  exists for normal dispersion 
($\gamma,\lambda$$>$0), and a family of dark solitons, $w_0=\sqrt{\lambda/
(4\gamma)}\,{\rm tanh}(\sqrt{|\lambda|}x){\rm e}^{i\lambda z}$, exists for 
anomalous dispersion ($\gamma,\lambda$$<$0).  With no SH, bright solitons, $w_0=\sqrt{2\lambda/|\gamma|}\,{\rm sech}
(\sqrt{2\lambda}x){\rm e}^{i\lambda z}$, exist for anomalous dispersion 
($\gamma$$<$0,$\lambda$$>$0), whereas dark solitons, $w_0=\sqrt{|\lambda|
/\gamma}{\rm tanh}(\sqrt{|\lambda|}x) {\rm e}^{i\lambda z}$, exist for 
normal dispersion ($\gamma$$>$0,$\lambda$$<$0).
We test these solutions by mapping them back to the variables $w$ and $v$ 
and launching them as initial conditions in simulations of the field 
Eqs.~(\ref{field_eqns}).  

\begin{figure}
  \centerline{\hbox{
  \psfig{figure=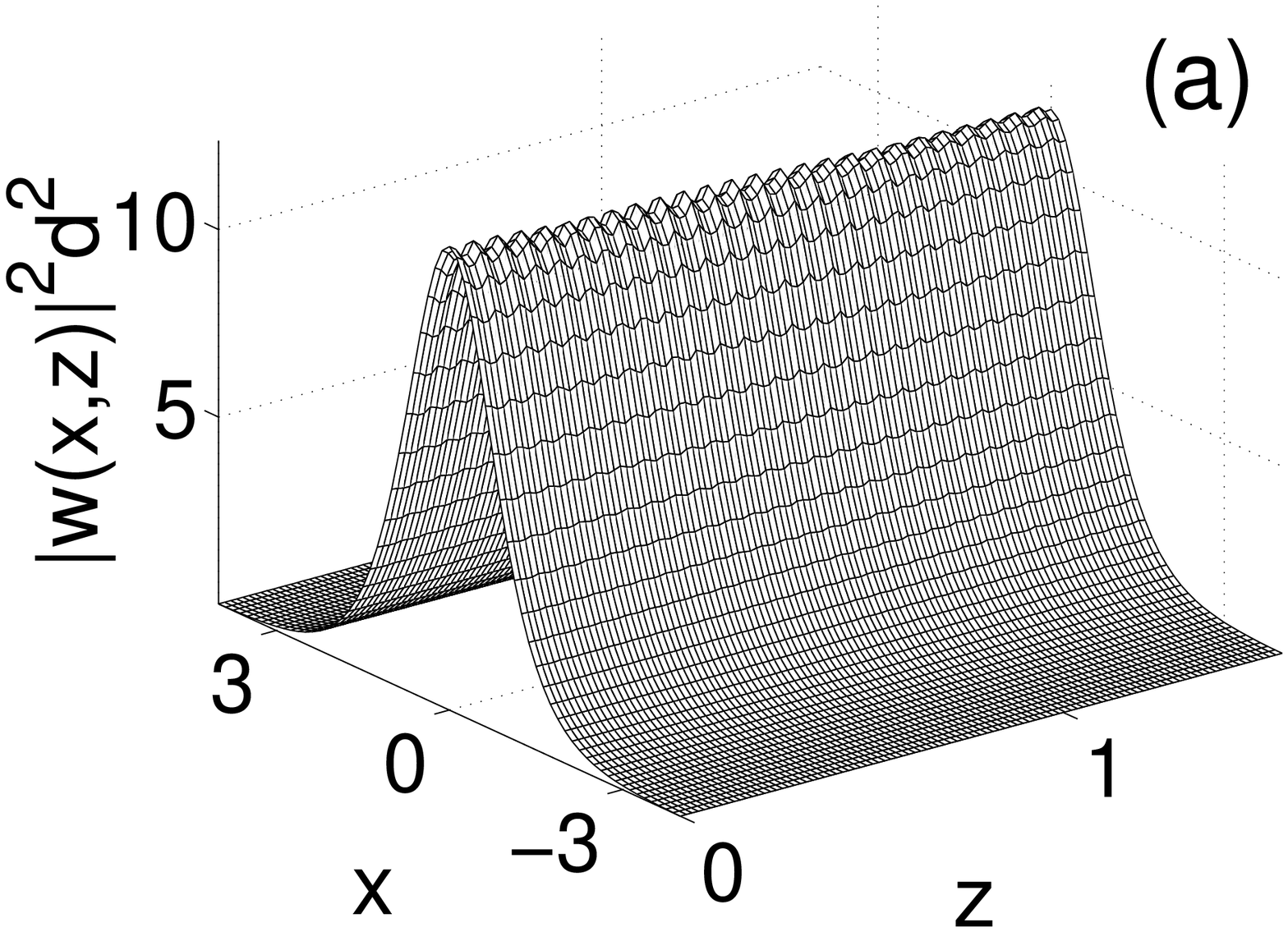,width=\figwidtha}
  \psfig{figure=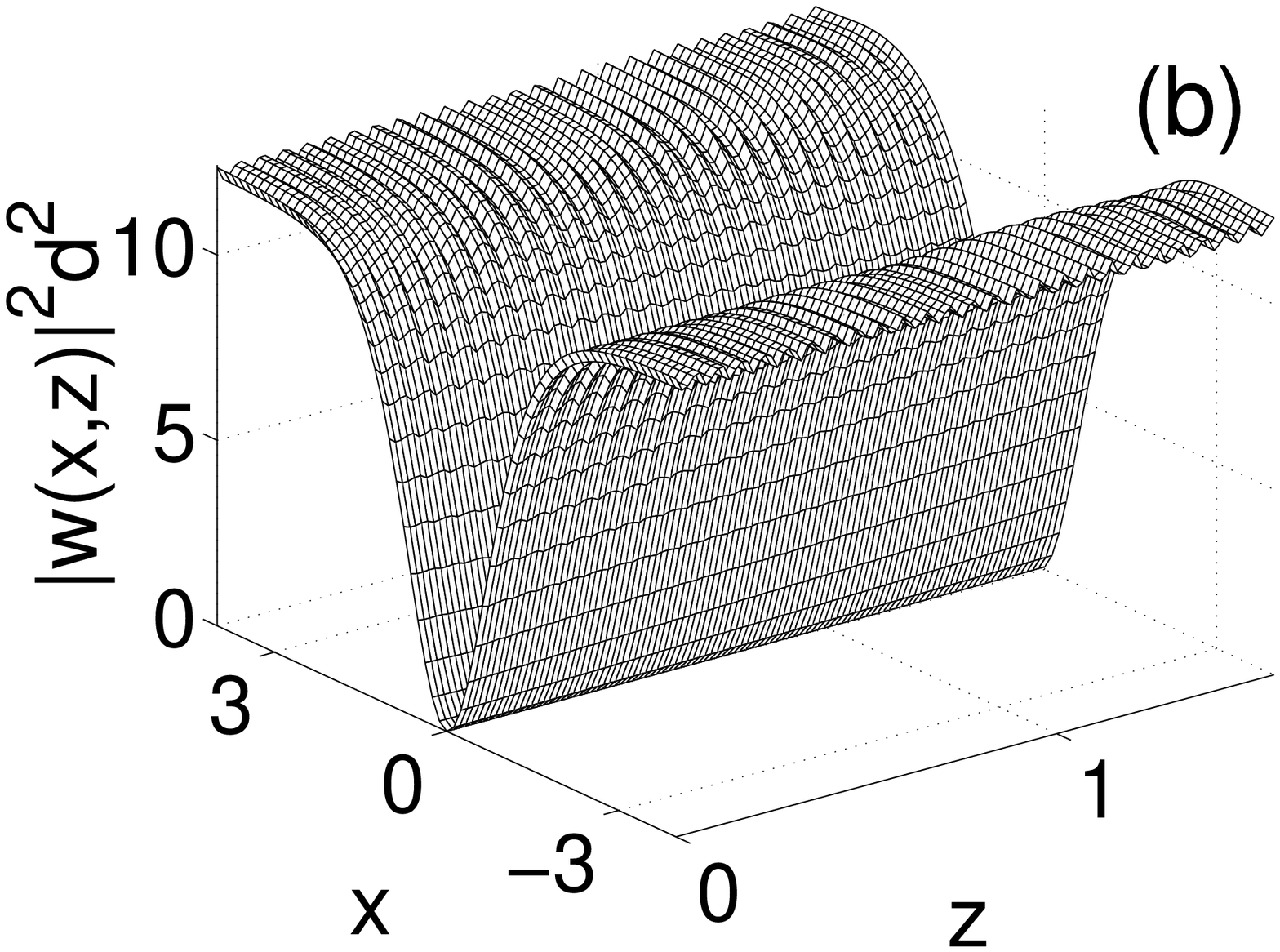,width=\figwidtha}}}
  \caption{(a) Bright and (b) dark solitons propagating in QNPCs with
  no effective $\chi^{(2)}$ nonlinearity. Shown is the scaled intensity of
  the fundamental for $|\kappa|$=100, $\tilde\beta$=0, and (a) $\gamma<0$,
  $\lambda$=1 and (b) $\gamma>0$, $\lambda$=$-1$. The SH is zero.} 
\label{darksoliton}
\end{figure}

Figure \ref{darksoliton} shows the evolution of a bright and dark soliton
with no SH over a distance of 25 grating periods.
The SH displays small, regular oscillations around the mean 
value zero, corresponding to the oscillations of the fundamental seen 
in Fig.~\ref{darksoliton}.
The simulations thus confirm that bright and dark solitons can indeed 
propagate in a stable manner in QNPCs with no effective $\chi^{(2)}$ 
nonlinearity. 

In summary, we have shown that bright solitons exist and propagate 
in a stable manner in 1D quadratic nonlinear photonic crystals (QNPCs) 
with many types and combinations of linear and nonlinear QPM gratings. 
By deriving first-order averaged equations, we have shown that such 
QNPCs have an induced cubic nonlinearity, and we have numerically found 
previously unknown families of bright solitons.
Even with no effective quadratic nonlinearity, the QNPCs support both 
bright and dark solitons due to the induced cubic nonlinearity. 
We have found analytical expressions for these solitons and shown that they
propagate in a stable manner.

Dark solitons are always unstable in homogeneous $\chi^{(2)}$ media in
settings for type I SHG, due to modulational instability of the back-ground 
plane-waves \cite{dark95}.
Our results show, for the first time, a dark soliton that appears to be stable
under propagation, with the stabilizing mechanism necessarily originating 
from the photonic crystal structure of the QNPC.
An analysis of dark solitons in the general case is being carried out.

The project is supported by the Danish Technical Research Council 
through Talent Grant No.~9800400.

\end{multicols}


\begin{thebibliography}{}

\bibitem{Sou96}
   NATO ASI Series: {\em Photonic Band Gap Materials}, ed.~C.M.~Soukoulis 
   (Kluwer Academic Publishers, 1996) 

\bibitem{Leo_etal00}
   S.W. Leonard {\em et al.},
   \prb {\bf 61}, R2389 (2000).

\bibitem{BusJoh99}
   K. Busch and S. John,
   \prl {\bf 83}, 967 (1999).

\bibitem{LarHibMizSte90}
   S.~Larochelle {\em et al}.,
   Electron.~Lett.~{\bf 26} (1990) 1459.

\bibitem{SteHagTor96}
   For an review see 
   G. Stegeman, D.J. Hagan, and L. Torner,
   Opt. Quantum Electron. {\bf 28}, 1691 (1996).

\bibitem{ArmBloDucPer62}
   J.A. Armstrong {\em et al.}, 
   Phys. Rev. {\bf 127}, 1918 (1962).

\bibitem{TanBey73}
   C.L. Tang and P.P. Bey, 
   {IEEE} J. Quantum Electron. {\bf 9}, 9 (1973).

\bibitem{JasArvLau86}
   B. Jaskorzynska, G. Arvidsson, and F. Laurell, 
   Spie: Integrated Optical Circuit Engineering {III} {\bf 251}, 221 (1986).

\bibitem{FejMagJunBye92}
   M.M. Fejer {\em et al.}, 
   {IEEE} J. Quantum Electron. {\bf 28}, 2631 (1992).

\bibitem{NPC_chi2_bragg}
   N.~Bloembergen and A.J.~Sievers,
   Appl.~Phys.~Lett.~{\bf 17} (1970) 483.

\bibitem{Tor98}
   For a review see 
   L. Torner, in {\em Beam Shaping and Control with Nonlinear Optics}, 
   eds. F. Kajzer and R. Reinisch (Plenum, New York, 1998).

\bibitem{gap}
   Yu.S. Kivshar, 
   \pre {\bf 51}, 1613 (1995). 
   For a recent overview see
   A. Arraf and C.M. de Sterke, 
   \pre {\bf 58}, 7951 (1998).

\bibitem{Hel00}
   A.A. Helmy {\em et al.},
   "Quasi-phase-matching in GAAS-ALAS superlattice waveguides via bandgap
   tuning using quantum well intermixing",
   \ol, {\em to appear}.

\bibitem{ClaBanKiv97}
   C.B. Clausen, O. Bang and Y.S. Kivshar, \prl {\bf 78}, 4749 (1997). 

\bibitem{DMsol}
   J.H.B. Nijhof {\em et al.},
   Electron. Lett. {\bf 33}, 1726 (1997).
   S. K. Turitsyn and E. G. Shapiro, 
   \ol {\bf 23}, 682 (1998)

\bibitem{MenSchTor94}
   C.R. Menyuk, R. Schiek, and L. Torner, \josab {\bf 11}, 2434 (1994).
   O. Bang, {\em ibid.} {\bf 14}, 51 (1997).

\bibitem{KobLedBanKiv98}
   A. Kobyakov {\em et al.},
   \ol {\bf 23} (1998) 506.

\bibitem{BanClaChrTor99}
   O. Bang {\em et al.}, 
   \ol {\bf 24}, 1413 (1999).


\bibitem{Pet96}
   D.V. Petrov, 
   \oc {\bf 131}, 102 (1996).

\bibitem{dark95}
   S. Trillo and P. Ferro, \ol {\bf 20}, 438 (1995).
   A.V. Buryak and Yu.S. Kivshar, \ol {\bf 20}, 834 (1995).


\end{thebibliography}
\end{document}